\documentclass[pra,twocolumn,showpacs,superscriptaddress,preprintnumbers,amsmath,amssymb,aps]{revtex4-1}
\usepackage{mathrsfs}
\usepackage{amsmath}
\usepackage{amssymb}
\usepackage{graphicx}
\usepackage{dcolumn}
\usepackage{bm}
\usepackage{subfigure}
\usepackage{color}
\usepackage{ifpdf}
\usepackage{epstopdf}
\usepackage{CJK}
\usepackage{verbatim}
\usepackage{bbm}
\usepackage{multirow}
\usepackage{booktabs}
\usepackage{color}
\usepackage[
pdfstartview=FitH,%
bookmarks=true,%
bookmarksopen=true,%
breaklinks=true,%
colorlinks=true,%
linkcolor=blue,anchorcolor=blue,%
citecolor=blue,filecolor=blue,%
menucolor=blue,pagecolor=blue,%
urlcolor=blue]{hyperref}
\usepackage{float}
\usepackage[toc,page,title,titletoc,header]{appendix}
\usepackage{ragged2e}
\usepackage{lipsum}

\usepackage{array}
\usepackage{booktabs}

\usepackage{color}

\def\be{\begin{equation}}
	\def\ee{\end{equation}}
\def\bea{\begin{eqnarray}}
	\def\eea{\end{eqnarray}}
\def\nn{\nonumber}

\begin{document}
\begin{CJK*}{GBK}{song}
		
\title{The dominant scattering channel induced by two-body collision of D-band atoms in a triangular optical lattice}
\author{Xinxin Guo}
\affiliation{State Key Laboratory of Advanced Optical Communication Systems and Networks, Department of Electronics, Peking University, Beijing 100871, China}
\author{Zhongcheng Yu}
\affiliation{State Key Laboratory of Advanced Optical Communication Systems and Networks, Department of Electronics, Peking University, Beijing 100871, China}
\author{Peng Peng}
\affiliation{State Key Laboratory of Advanced Optical Communication Systems and Networks, Department of Electronics, Peking University, Beijing 100871, China}
\author{Guoling Yin}
\affiliation{State Key Laboratory of Quantum Optics and Quantum Optics Devices, Institute of Opto-Electronics, Shanxi University, Taiyuan 030006, China}
\affiliation{Collaborative Innovation Center of Extreme Optics, Shanxi University, Taiyuan, Shanxi 030006, China}
\author{Shengjie Jin}
\affiliation{State Key Laboratory of Advanced Optical Communication Systems and Networks, Department of Electronics, Peking University, Beijing 100871, China}
\author{Xuzong Chen}
\affiliation{State Key Laboratory of Advanced Optical Communication Systems and Networks, Department of Electronics, Peking University, Beijing 100871, China}
\author{Xiaoji Zhou}\email{xjzhou@pku.edu.cn}
\affiliation{State Key Laboratory of Advanced Optical Communication Systems and Networks, Department of Electronics, Peking University, Beijing 100871, China}
\affiliation{Collaborative Innovation Center of Extreme Optics, Shanxi University, Taiyuan, Shanxi 030006, China}
\date{\today}

\begin{abstract}
The mechanism of atomic collisions in excited bands plays an important role both in the atomic dynamics in high bands of optical lattices and simulation of condensed matter physics. Atoms distributed in an excited band of an optical lattice can collide and decay to other bands through different scattering channels. In the excited bands of a one dimensional lattice there is
no significant difference between the cross sections to different scattering channels, due to the sameness of all of the geometrical
couplings. Here, we investigate the collisional scattering channels for atoms in the excited bands of a triangular optical lattice and demonstrate a dominant scattering channel in the experiment. A shortcut method is utilized to load Bose-Einstein condensate of  $^{87} {\rm Rb}$ atoms into $\Gamma$ point of the first D band with zero quasi-momentum in the triangular optical lattice. After some evolution time, the number of atoms scattering into the S band induced by two-body collisions is around four times the number that scatter into the second most populated band. Our numerical calculation shows that the $ss$ scattering channel is dominant, which is roughly consistent with the experimental measurement. The appearance of dominant scattering channels in a triangular optical lattice is owing to non-orthogonal lattice vectors. This work is helpful for the research on many-body systems and directional enhancement in optical lattices.
\end{abstract}
		
		
\maketitle
\end{CJK*}
	
\section{Introduction}

In many-body systems, collision is one of the most important interactions. The investigations of low energy collisions in atomic \cite{PhysRevLett.81.938,PhysRevLett.91.163201,RN1701,PhysRevA.93.022705,PhysRevLett.103.163201,RevModPhys.71.1,Kemper_1984,PhysRevA.79.050701}, ionic \cite{PhysRevLett.109.233202,PhysRevA.95.032709}, and electronic systems \cite{SEITOV202026,RN1703,RN1704} have been a subject of intense research in recent years.
As for ultracold atoms, the collision rate is one of the major factors to determine the coherence time of the system \cite{PhysRevA.55.R2511,RevModPhys.82.1225}. The scattering cross section is defined to describe the collision rate, which 
has been extensively studied experimentally and theoretically \cite{PhysRevA.81.012713,PhysRevLett.97.180404}.

The research on ultracold atoms in optical lattices has attracted much attention for their abundant properties especially in excited bands of lattice, including dynamical superfluidity in higher lattice orbitals \cite{PhysRevLett.121.265301,PhysRevLett.126.035301,PhysRevLett.99.200405,PhysRevLett.106.015302,PhysRevA.94.033624}, staggered orbital currents\cite{PhysRevA.74.013607} and decay mechanism in excited bands\cite{PhysRevA.87.063638,ramsey}.
The collision in optical lattices not only changes the internal state of atoms \cite{PhysRevA.81.023605,PhysRevA.101.013636} but also influences the external state \cite{PhysRevA.87.063638}. For instance, two atoms on an excited band in the optical lattice would jump to other bands owing to two-body collision \cite{PhysRevA.90.013602}. The path for atoms scattering from a certain initial state to a final state is defined as a scattering channel.	
So far, several theoretical and experimental work has been achieved to investigate the effect of atomic collisions in one, two, and mixed-dimensional optical lattices \cite{PhysRevA.72.053604,RN1698,PhysRevA.88.033615,PhysRevLett.104.153202,PhysRevLett.111.205302}, such as the measurement of collision rate for atoms in the P band \cite{PhysRevLett.111.205302} and the observation of scattering halos \cite{RN1697}, etc.  
Recently, we have demonstrated the cross section of excited bands in one dimensional (1D) lattice experimentally\cite{arXiv:2104.08794}, where the cross sections to each band have no significant difference and no dominant scattering channel exists.
Different from the situation in the 1D lattice, two dimensional (2D) lattice has more information on geometry and dimension, whereas the study of scattering channels in 2D optical lattice has remained unexplored systematically.

Here, we perform theoretical and experimental studies of scattering channels induced by two-body collisions at $\Gamma$ point of the first D band ($\rm D_1$ band) in a triangular optical lattice and demonstrate a dominant scattering channel. 
Our experiment starts from a Bose-Einstein condensate (BEC) in a harmonic trap, and then we use the shortcut method \cite{Zhou_2018} to load the atoms into the $\Gamma$ point of $\rm D_1$ band in the triangular optical lattice. After holding the atoms for a certain time in the optical lattice, we apply band mapping technique to get the distribution of atoms in reciprocal space \cite{PhysRevLett.99.200405,PhysRevLett.94.080403}.
We quantitatively measure the number of atoms in different bands through the absorption images obtained after time of flight.
$55.8\%$ atoms jump to the S band (the first Brillouin Zone(BZ)), while only about $10\%$ atoms jump to the two P bands (2nd, 3rd BZ) respectively and about $10\%$ atoms remain in the $\rm D_1$ band (4th BZ).
Meanwhile, theoretical calculation indicates that the scattering channel where two atoms jump from the $\rm D_1$ band to the S band is dominant. By adding up all the scattering channels to the same final state, we get that the cross section to the S band is $57.3\%$ of the total cross section, agreeing with experimental results. 
The reason for the dominant channel may be non-orthogonal lattice vectors which produce a term of potential both affected by position x and y, and decreases the overlapping area of eigenstates between bands with different parity.
This work contributes to the control of external states of atoms in an optical lattice, and the dominant scattering channel is possibly used for realizing directional enhancement.

In Sec. II, we describe the experimental process in the triangular lattice. Sec. III
introduces the collision model and scattering channels in 2D lattice. Then we calculate the cross section of scattering channels in the square and triangular optical lattice, respectively.  
In Sec. IV, we demonstrate the experimental result and give the normalized scattering cross section of each band. Then we compare the experiments with theoretical calculations.
In Sec. V, we compare triangular lattice with bipartite lattices, and analyze the connection between lattice geometry and the dominant scattering channel. Finally, we give a conclusion in Sec. VI.

\section{Experimental description}
\begin{figure}
\includegraphics[width=0.45\textwidth]{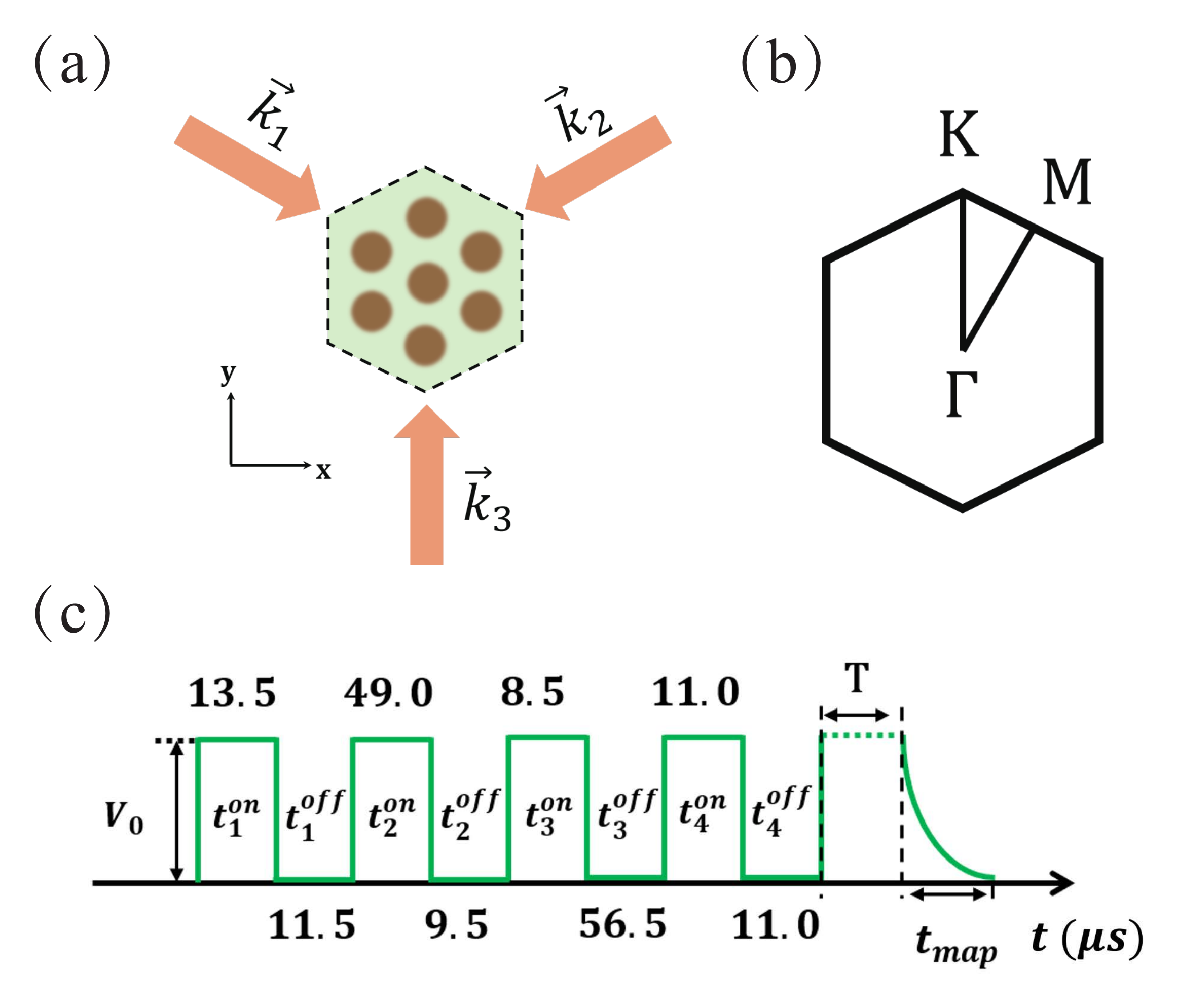}
\caption
{(a) gives the diagram of a triangular optical lattice. The arrows represent the laser beams with wave vectors $\vec{k_i}$, where $i=1,2,3$ and $\vec{k_3}=-(\vec{k_1}+\vec{k_2})$.
(b) is the reduced Brillouin zone of the triangular lattice corresponding to (a). The high symmetry line of the band $\rm K-\Gamma-M-K$ is marked.
(c) The time sequence diagram of lattice depth for loading atoms into the $\Gamma$ point of $\rm D_1$ band in triangular lattice. The four pulses $/t^{\rm on}_i/t^{\rm off}_i/$, $i=1,2,3,4$ form a shortcut sequence. After an evolution time $T$, the lattice beam intensity decreases to zero adiabatically in $t_{map}=1 ~{\rm ms}$. The lattice depth of the time sequence is $V_{0}=3.0 ~{\rm E_r}$.
}\label{fig:fig1_new}
\end{figure}

Our experiment is carried out in a 2D triangular optical lattice with tube-shaped lattice sites\cite{PhysRevLett.126.035301,Becker_2010,Guo:19,Jin_2019}. As shown in Fig. \ref{fig:fig1_new}(a), the triangular optical lattice is formed by three intersecting $\lambda=1064 ~{\rm nm}$ laser beams, which are linearly polarized perpendicular to the lattice plane (x-y plane). $\vec{k_1}$, $\vec{k_2}$ and $\vec{k_3}$ are wave vectors of the three laser beams with $120^\circ$ enclosing angles.
In the direction perpendicular to the lattice plane, atoms are weakly confined by an approximately harmonic potential. Fig. \ref{fig:fig1_new}(b) shows the reduced Brillouin Zone corresponding to the triangular lattice in Fig. \ref{fig:fig1_new}(a), and it marks the high symmetry line $\rm K-\Gamma-M-K$.

We start with a BEC of about $3\times 10^5$ atoms in the $\left|F=2,m_F=+2\right>$ state, which is confined in a hybrid trap with the harmonic trapping frequencies $(\omega_x,\omega_y,\omega_z)=2\pi\times(28,55,60)~{\rm Hz}$. Next, a nonadiabatic shortcut method is utilized to load BEC from the harmonic trap into the $\Gamma$ point (the quasi-momentum $\vec{q}=0$) of $\rm D_1$ band in the triangular optical lattice \cite{Zhou_2018,PhysRevLett.121.265301,PhysRevA.87.063638}. 
The duration and interval time sequence of  the optical pulses for shortcut is optimized to reach the target state with high fidelity. 
For the lattice depth $V_{0}=3.0 ~{\rm E_r}$, after optimizing, we get a four-pulse sequence as shown in Fig.\ref{fig:fig1_new} (c), and the on/off time of the lattice is $13.5/11.5/49.0/9.5/8.5/56.5/11.0/11.0 ~{\rm \mu s}$. The theoretical fidelity of the sequence can reach $99.95\%$ (More details in Appendix B).

After being loaded into the $\rm D_1$ band, the BEC in the optical lattice evolves for a certain time $T$. Then, we apply band mapping\cite{PhysRevLett.99.200405,PhysRevA.73.020702} by switching off the lattice potential adiabatically in the form $e^{-t_{map}/\tau}$, where the time constant $\tau=200~\rm{\mu s}$ for the total time $t_{\rm map}=1 ~\rm{ms}$, as shown in Fig.\ref{fig:fig1_new} (c). Atoms populated in the $n$th band with quasimomentum $\vec{q}$ and energy $E$ can be mapped to some point of the $n$th Brillouin zone with quasimomentum $\vec{q}$.
Finally, we take absorption imaging with time of flight(TOF) $t_{TOF}= 30~\rm{ms}$ to measure the quasi-momentum space distribution of atoms in each band.

\begin{figure*}[htp]
	\includegraphics[width=0.9\textwidth]{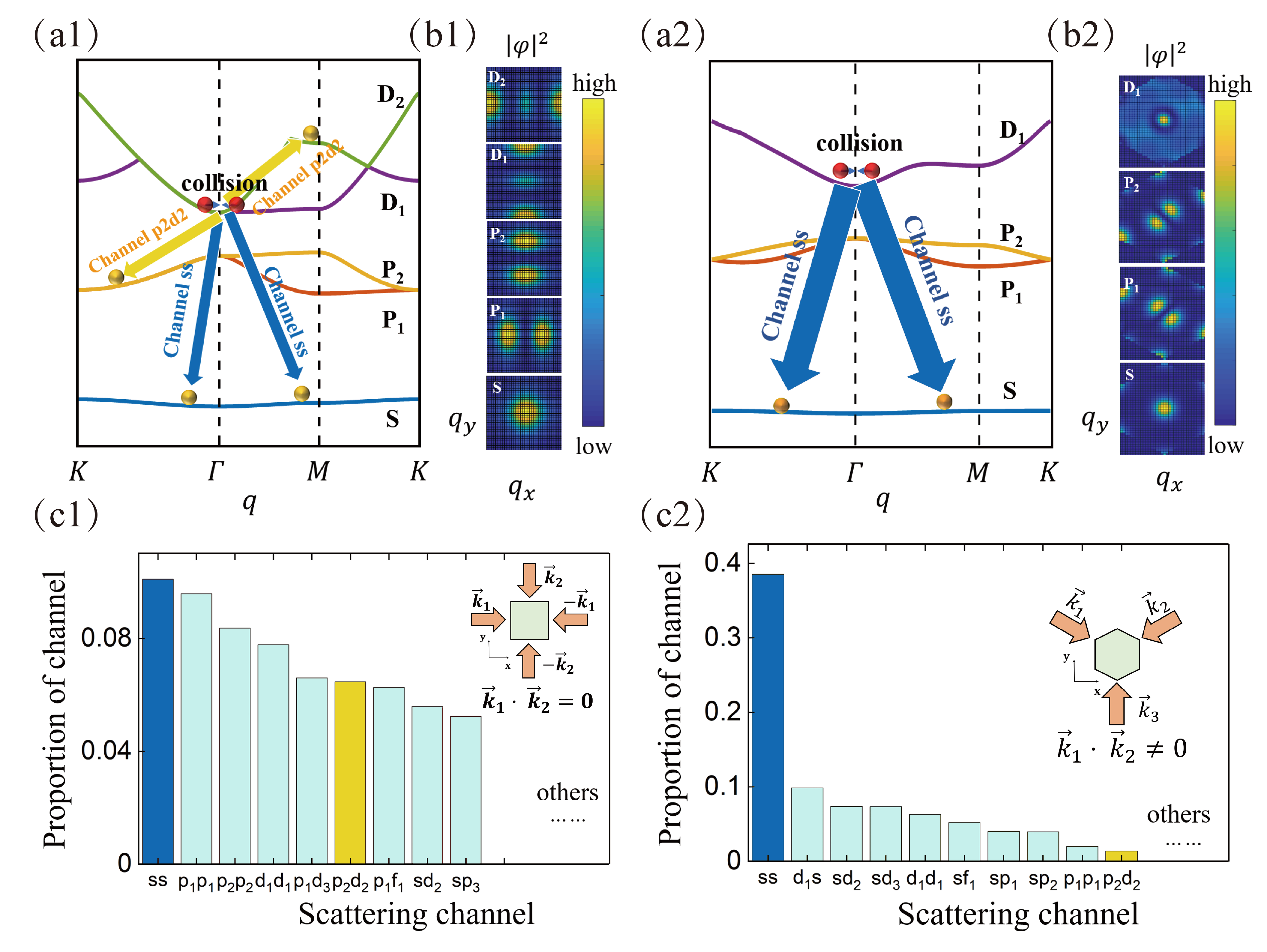}
	\caption{\textbf{Scattering channels of atoms in 2D lattice.}
		(a1) and (a2) show the band structure along the high symmetry lines in square and triangular lattice, respectively. The red and yellow spheres represent the atomic initial states and final states during the collision scattering. Arrows with the same color represent the scattering channel generated by a collision. Here we only draw the first few bands and scattering channels, and others are not in the figure.
		(b1) and (b2) show the squared modulus of the wave function $| u(\vec{q})|^2$ for the states at the $\Gamma$ points corresponding to bands in (a1) and (a2), respectively.
		The proportions of several main scattering channels are shown in (c1) (square lattice) and (c2) (triangular lattice). The inserts show the diagram of a square optical lattice and triangular optical lattice.
	}\label{fig:fig2_new}
\end{figure*}

\section{Collisional Scattering Process and Calculation of Scattering Channels}
In the above section, we describe the experimental process, and load the atoms into the $\Gamma$ point of $\rm D_1$ band. 
To study the evolution of atoms, in this section, we discuss the collisional scattering process of atoms in the excited bands of 2D optical lattice and calculate the cross section of the collisional scattering.

\subsection{Collisional Scattering Process}
The potential of 2D lattice $V_{\rm lattice}$ can be expressed as:
\begin{eqnarray}
\label{V}
&&V_{\rm lattice}=V_x \cos(\vec{k_x}\cdot \vec{r})+V_y \cos(\vec{k_y}\cdot \vec{r}) \\ \nn
&&+\sum_{a,b}V_{ab} \cos((a\vec{k_x}+b\vec{k_y})\cdot \vec{r}),
\end{eqnarray}
where $\vec{k}_x$ and $\vec{k}_y$ are the lattice vectors in x and y direction respectively, $\vec{r}$ is the position vector. $V_x$, $V_y$ and $V_{ab}$ are potential energy components in x, y and oblique directions, where $a,b$ are any non-zero integers. 
	
The first two terms on the right side of Eq.(\ref{V}) are independent whereas the last term is related to both the vector $\bf x$ and $\bf y$, defined as x-part, y-part and x-y dimensional coupling part correspondingly. For a 2D lattice,
the appearance of dimensional coupling part is due to the non-orthogonality of lattice vectors.
When $V_{ab}=0$, the lattice potential is independent in x and y direction, such as the square lattice. Comparably, when $V_{ab} \ne 0$, the lattice potential is not independent in x and y direction, like the triangular optical lattice.

Fig. \ref{fig:fig2_new} (a1) and (a2) show band structure along the high symmetry line $\rm K-\Gamma-M-K$ of square and triangular optical lattice, respectively. For a 2D optical lattice, there are one S band, two P bands ($\rm P_1$, $\rm P_2$) and four D bands ($\rm D_1$, $\rm D_2$, $\rm D_3$, $\rm D_4$).
For $^{87}$Rb BEC in optical lattice, the collisions are mainly low energy scattering, and the $s$-wave approximation is reasonable \cite{RN1697}. Further, during the measuring time (about several ${\rm ms}$), three-body collision could be neglected, of which characteristic time is several second in $^{87}$Rb BEC. In the following, we just consider the two-body $s$-wave collision, and assume that the atoms only undergo one collision during the scattering process.

As shown in Fig. \ref{fig:fig2_new}(a1) and (a2), atoms initially staying at the $\rm \Gamma$ point of $\rm D_1$ band would jump to other bands because of collisions, where the red and yellow spheres correspond to the initial and final states of the two atoms. As shown by the blue arrows in Fig. \ref{fig:fig2_new}(a1) and (a2), the two atoms both jump to S band, and we mark this case as $ss$ scattering channel. Similarly, if one of the two atoms jumps to $\rm P_2$ band and another one to the $\rm D_2$ band shown as the yellow arrows, it is called $p_2d_2$ scattering channel. Different choices of final states are defined as different scattering channels. Briefly, we only draw a few typical scattering channels, and in reality the atoms are possible to jump to any band. However, the scattering probability to each band is different, and the strength of scattering probability is defined as scattering cross section.

\subsection{Calculation of Scattering Channels}
In order to study the difference of the scattering process between square lattice and triangular lattice, we use the scattering theory to calculate the cross section of each scattering channel in those two types of lattices.
Two-body collisional scattering cross section for two atoms initially at the $\Gamma$ point($(q_x,q_y)$=(0,0)) of $\rm D_1$ band jumping to band $n_1$ and $n_2$  can be written as \cite{PhysRevA.87.063638} (more details in Appendix A):
\begin{eqnarray}
	\label{sigma_s}
	&&\sigma(n_1 ,n_2)=\\
	&&  \frac{4\pi m \hbar}{v_{a}}\int d\vec{q} \times |-2\pi i \frac{4\pi a_s}{m} \zeta_{n_1 ,n_2}(0,0;\vec{q},-\vec{q})|^2, \nn
\end{eqnarray}
where $v_{a}$ is the atomic velocity, $m$ is the atomic mass and $a_s$ is atomic $s$-wave scattering length. And the overlapping integral of eigenstates $\zeta_{n_1 ,n_2}(0,0;\vec{q},-\vec{q})$ is given by:
\begin{eqnarray}
	\label{Gamma_s}
	&&\zeta_{n_1 ,n_2}(0,0;\vec{q},-\vec{q})=\int d\vec{r} \\ \nn
	&&\times u^*_{n_1,\vec{q}}(\vec{r}) u^*_{n_2,-\vec{q}}(\vec{r}) u_{d,0}(\vec{r}) u_{d,0}(\vec{r})
\end{eqnarray} 
where $u_{n_i,\vec{q}}$ (i=1,2...) is the eigenstate at quasi-momentum $\vec{q}$ in band $n_i$. In the calculation, we assume the periodic boundary conditions, and consider that $|\zeta_{n_1 ,n_2}(0,0;\vec{q},-\vec{q})|^2=\int d\vec{r}\times |u^*_{n_1,\vec{q}}(\vec{r}) u^*_{n_2,-\vec{q}}(\vec{r}) u_{d,0}(\vec{r}) u_{d,0}(\vec{r})|^2$. Fig. \ref{fig:fig2_new} (b1) and (b2) show modulus square of the eigenstates $u_{n_i,\vec{q}}$ at the $\Gamma$ point of each band in square and triangular optical lattice, which are calculated by secular equations of the optical lattice, where we choose the wavelength of optical lattice $\lambda=1064 ~{\rm nm}$ and lattice depth $V_0=3 ~{\rm E_r}$ (${\rm E_r}=\frac{h^2}{2m\lambda^2}$ is the single-photon recoil energy).

To study the proportion of each scattering channel, we consider the lowest seven bands of the square and triangular optical lattice, because the scattering channels of higher bands are weak. Then we calculate the scattering channels, as shown in Fig. \ref{fig:fig2_new} (c1) and (c2). In the square lattice, the cross section of the strongest channel $ss$, $p_1p_1$ and $p_2p_2$ are all around $10\%$ of the total cross section respectively. Besides, there are many other smaller channels included in 'Others'. There is no significant difference in scattering cross section values among the first six channels, which means that there is no dominant scattering channel in square lattice.
By contrast, in the triangular lattice, the proportion of scattering cross section of the $ss$ channel is $38.5\%$, while that of the second strong channel $d_1s$ is only $9.8\%$.
Besides, the proportion of other channels are much lower than that of channel $ss$. Consequently, the channel $ss$ is dominant in the two body scattering process of triangular optical lattice.

\begin{figure}
	\includegraphics[width=0.5\textwidth]{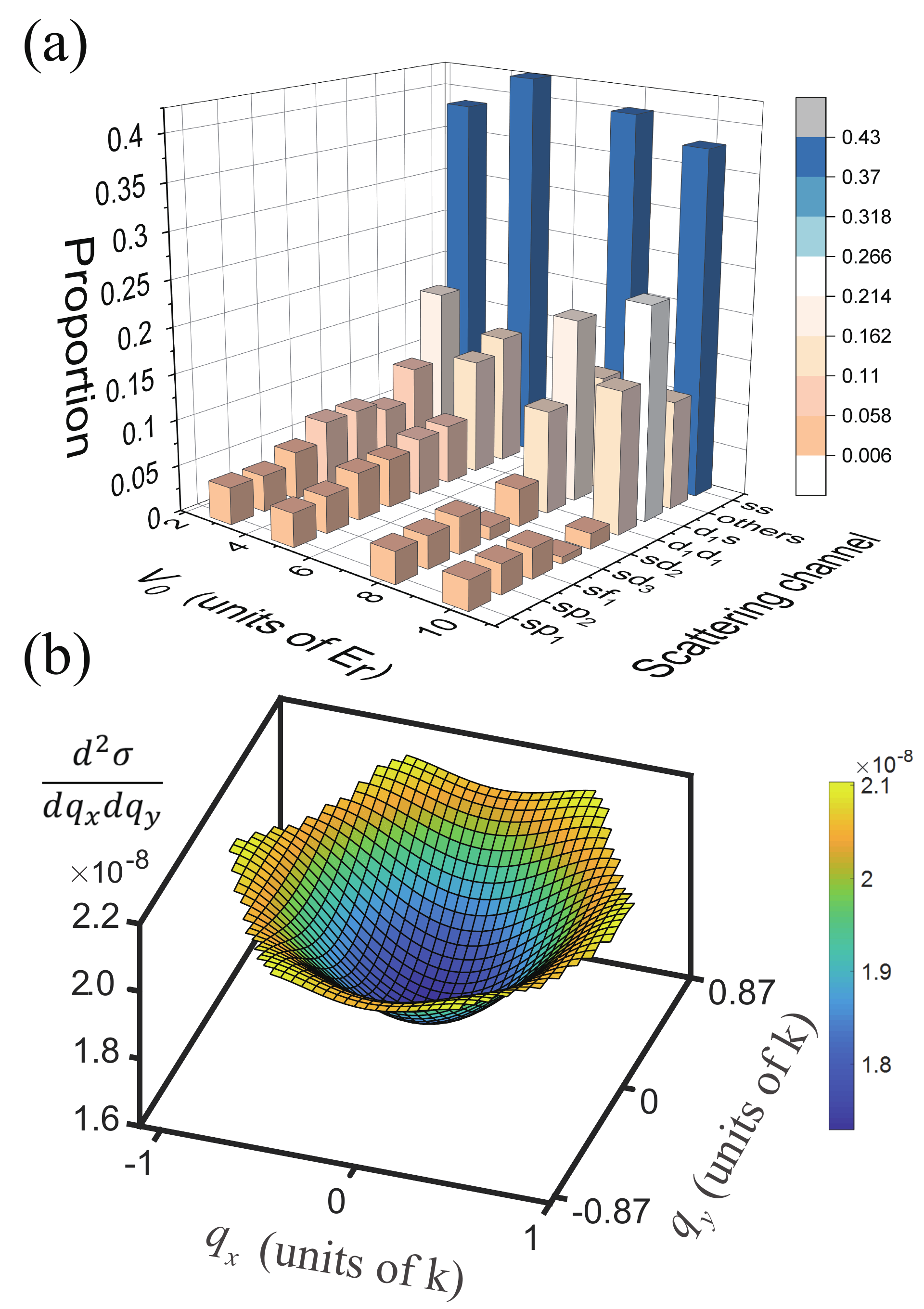}
	\caption
	{(a) The proportion of each scattering channel for different lattice depths $V_0=$ 3, 5, 8, and 10 ${\rm E_r}$. Only several main channels are shown in the figure, and the sum of the proportion of others is shown in 'others'.
	(b) Normalized differential cross section of scattering channel $ss$ with $(q_x,q_y)$ in quasi-momentum space, where $k=\frac{2\pi}{\lambda}$.
	}\label{fig:fig1.2}
\end{figure}

Further, we study the influence of the lattice depth $V_{0}$ on the dominant channel $ss$ in the triangular lattice. Using the same method, we calculate the normalized scattering cross sections of each channel with different lattice depth $V_{0}$, as shown in Fig. \ref{fig:fig1.2}(a).
Among those lattice depths, the scattering channel $ss$ is always the dominant scattering channel.

Besides, Eq.(\ref{sigma_s}) can give the differential cross section $\frac{{\rm d}^2\sigma}{{\rm d}\vec{q}}(\vec{q})$ of scattering channel, which denotes the scattering probability to different quasi-momentum. Fig.\ref{fig:fig1.2}(b) shows the normalized differential cross section of the channel $ss$ in triangular optical lattice at $V_{0}=3 ~{\rm E_r}$. The differential cross section covers the entire Brillouin zone, and the value of differential cross section at the center of the Brillouin zone is $4\%$ lower than that at the edge.
Since this difference is small, the atoms would approximately uniformly scatter to S band, and the $1{\rm st}$ BZ almost evenly filled by atoms should be observed.

\section{Experimental results and analysis}
	
\subsection{Experimental results}

\begin{figure}
	\includegraphics[width=0.5\textwidth]{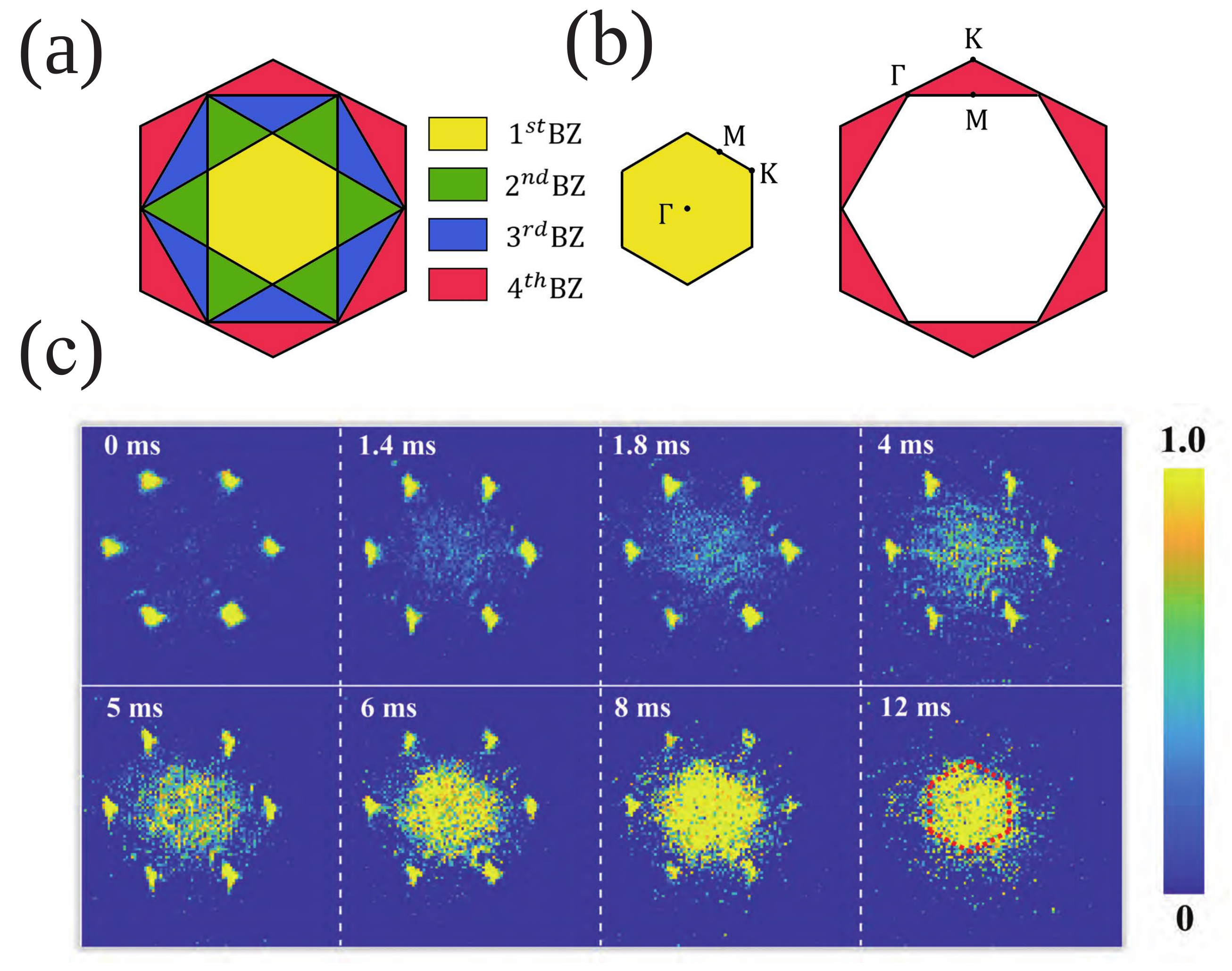}
	\caption{
		(a) Schematic diagram of the first four Brillouin zones of a triangular optical lattice. Yellow, green, blue, red areas, represent $1st, 2nd, 3rd, 4th$ BZ, respectively. 
		(b) The position of high symmetry point in the $1st$ and $4th$ extended Brillouin zone. The yellow area represents the $1st$ BZ, and red area represents the $4th$ BZ. The corresponding positions of point $K, \Gamma, M$ in the two areas are marked.
		(c) Observation of the atomic distribution over different evolution time. A given color represents the same number of atoms in each panel, where the maximum number of atoms is normalized to 1. The red box marks the $1st$ BZ. 
	}\label{fig:fig3.2}
\end{figure}

In the experiment, the distribution of atoms in quasi-momentum space could be observed after band mapping. The atomic distribution in the $n$th band is mapped to the area within the $n$th Brillouin zone in extended Brillouin zones. Fig. \ref{fig:fig3.2}(a) shows the first four extended Brillouin zones of the triangular lattice, yellow, green, blue, and red areas, corresponding to the S, $\rm P_1$, $\rm P_2$, and $\rm D_1$ bands, respectively. And Fig. \ref{fig:fig3.2}(b) shows the points with high symmetry $\rm \Gamma$, $\rm K$ and $\rm M$ in the $1st$ and $4th$ extended Brillouin zone .

\begin{figure}
	\includegraphics[width=0.45\textwidth]{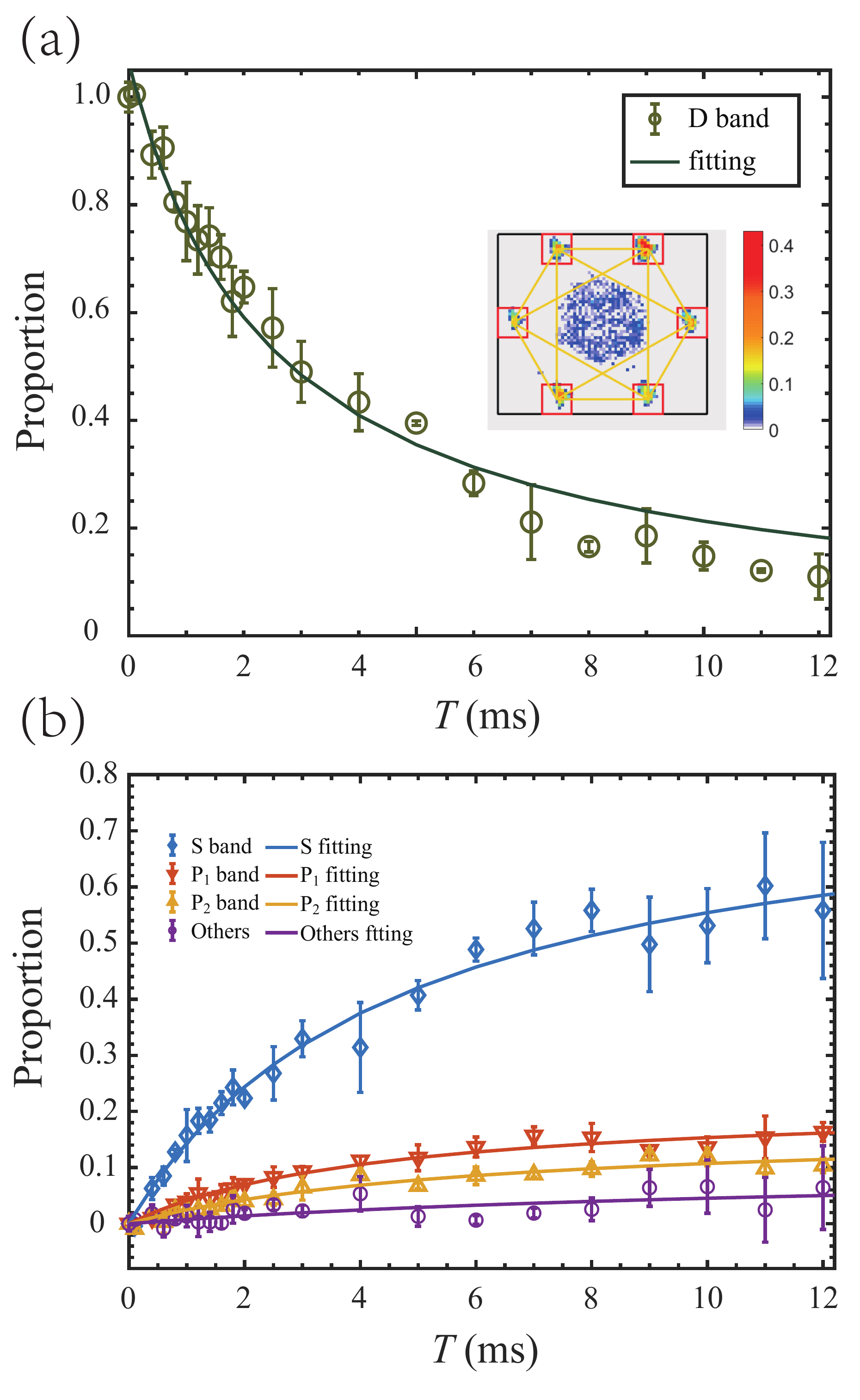}
	\caption{\textbf{The normalized proportion of atoms over different evolution time.} 
		The proportion of $\rm D_1$ band over the evolution time is shown in (a), and the proportion of other band is shown in (b). The green circles in panel (a) denote the atomic proportion in $\rm D_1$ band and the blue diamonds (orange down triangles, yellow up triangles, and purple circles) in panel (b) represent the atomic proportion in S ($\rm P_1$,$\rm P_2$ band and others), of which the solid lines with the same color are fitting lines.
		The insert in (a) show the method to extract atoms numbers for different Brillouin zones. The error bars represent the standard deviation of the five times measurement. The lattice depth is $ \rm{V_{0}=3 ~E_r}$.
	}\label{fig:fig3}
\end{figure}

Fig.\ref{fig:fig3.2}(c) shows the band population of ultra-cold atoms, which initially stay at $\Gamma$ point of $\rm D_1$ band, versus different evolution time $T$ in the experiment. When $T=0 ~{\rm ms}$, the atoms almost distribute at six points in the fourth Brillouin zone, where is the $\Gamma$ point of $\rm D_1$ band. As evolution time $T$ increases, the atoms gradually scatter to other bands. At $T= 1.8 ~{\rm ms}$, a considerable number of atoms could be observed in the $1{\rm st}$ BZ, while atoms populating in other BZs are few. When $T= 4.0 ~{\rm ms}$, the number of atoms at $1{\rm st}$ BZ and the six points of $4{\rm th}$ BZ are close. At $T= 8.0 ~{\rm ms}$, a prominent part of atoms at $\Gamma$ point of $\rm D_1$ band decay, and the atoms at $1{\rm st}$ BZ are more than those in any other areas obviously. Finally, at $T= 12.0 ~{\rm ms}$, atoms distributed at the $\Gamma$ point of $\rm D_1$ band completely decay, and the $1{\rm st}$ BZ is nearly tiled with atoms.
The population of atoms in the first Brillouin zone forms a hexagon with a clear outline, as the red box marks, agreeing with the calculation of differential cross section shown in Fig. \ref{fig:fig1.2}(b).

\subsection{Comparision between experiments and theoretical calculation}

	\begin{figure}
		\includegraphics[width=0.45\textwidth]{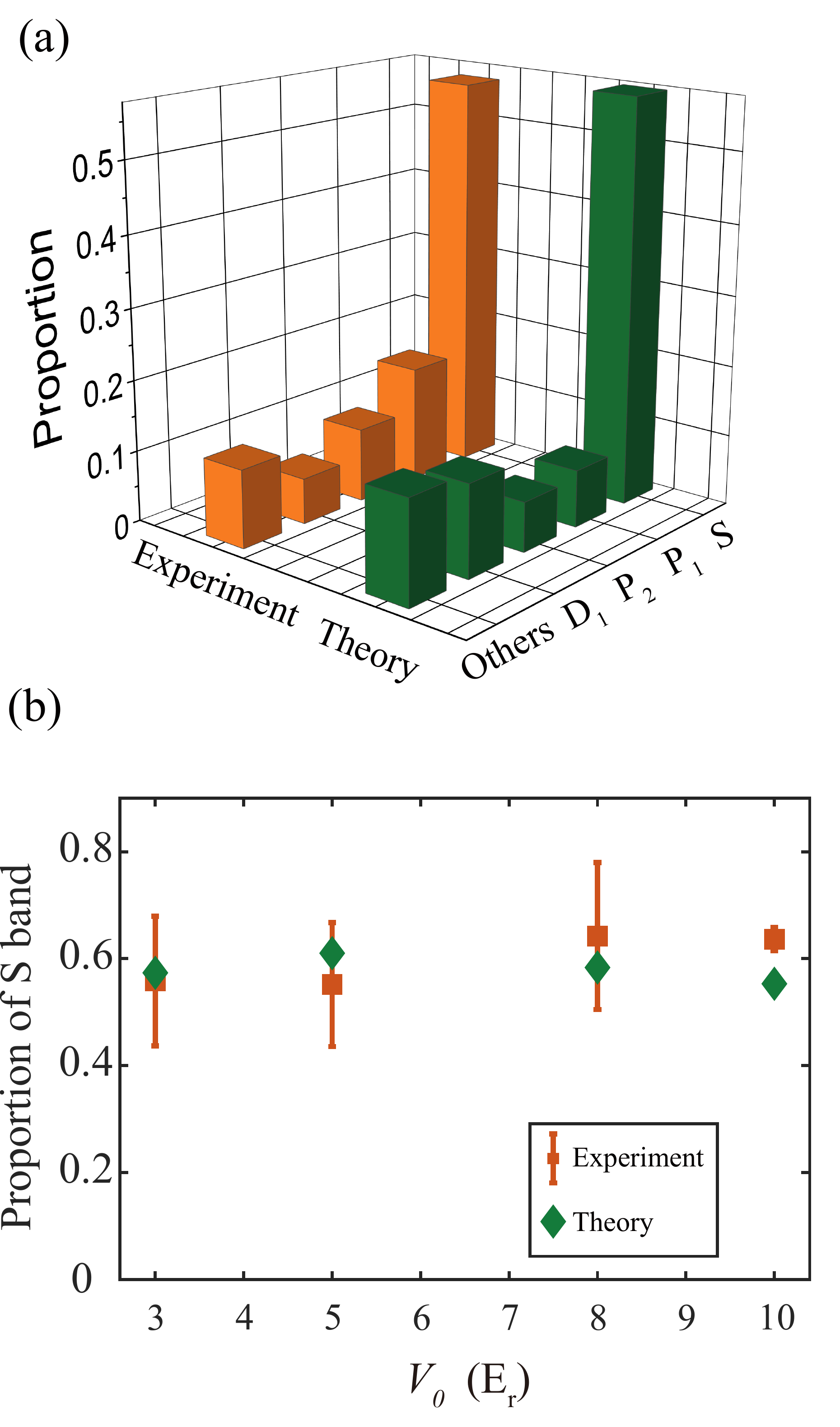}
		\caption{(a) The proportion of atoms in the S, $\rm P_1$, $\rm P_2$, $\rm D_1$ and other bands, when atoms initially at $\Gamma$ point of $\rm D_1$ band all decay. The experimental points (orange bar) are consistent with theoretical points (green bar). (b) The proportion of S band for different lattice depths $V_0$. Experiment (Theory) points are denoted by orange squares (green diamonds). The error bars represent the standard deviation of the five times measurement.
		}\label{fig:fig5}
	\end{figure}
	
To quantitatively compare the theoretical and experimental scattering cross sections, we calculate the proportion of atoms in each Brillouin zone of Fig. \ref{fig:fig3.2}(c). 
We define the proportion of atoms in each Brillouin zone as:
\begin{eqnarray}
	\label{def_proportion}
	p_{i}=N_{i}/N_{0},
\end{eqnarray} 
where ${i}={\rm S,P_1,P_2,D_1}$, and $N_i$ is the number of atoms in band $i$. $N_0$ is the total number of atoms.

The proportion of atoms corresponding to Fig. \ref{fig:fig3.2}(c) is shown in Fig.\ref{fig:fig3}, where Fig.\ref{fig:fig3} (a) shows the proportion of atoms in $\rm D_1$ band and Fig.\ref{fig:fig3} (b) shows the proportion of S, $\rm P_1$, $\rm P_2$ and other higher bands.	The error bar of each data represents the standard deviation of five times measurement.
The insert of Fig. \ref{fig:fig3} (a) shows our division of atomic distribution. The six red rectangles mark the $\Gamma$ point of $\rm D_1$ band, and the number of atoms in those areas is considered as $N_{D_1}$. The yellow lines mark the $1{\rm st}, 2{\rm nd}, 3{\rm rd}$ BZs, and the numbers of atoms in these three BZs (except the area in the six red rectangles) are defined as $N_{S},N_{P_1},N_{P_2}$ respectively. We use the six red rectangles instead of the $4^{th}$ BZ to make the initial atomic count of $\rm D_1$ band more accurate, but it will make the count of atoms at the end of evolution smaller.

The green hollow circles in Fig.\ref{fig:fig3} (a) show the proportion of atoms at $\Gamma$ point of $\rm D_1$ band. In Fig. \ref{fig:fig3} (b), the blue (orange, yellow) points indicate the proportion of atoms in S ($P_1$, $P_2$) band, and the purple points reveal the atoms in other higher bands. The number of atoms on $\rm D_1$ band at $T=0 ~{\rm ms}$ is defined as the total atom number. The solid lines fit the experimental points by functions $n_0(1-1/(1+KT))$ \cite{PhysRevA.87.063638}, where $n_0$ and $K$ are fitted parameters.	
At first, the atoms are loaded into the $\Gamma$ point of $\rm D_1$ band. 
With time increasing, the atoms at the $\Gamma$ point of $\rm D_1$ band gradually scatter to other bands. Hence, the proportion $p_{\rm D_1}$ reduces, while the proportion of atoms in other bands increases. Whereas, the increase rate of $p_s$ is much faster than that of $\rm P_1$, $\rm P_2$ band and others. At $T=5 ~{\rm ms}$ the proportion $p_{\rm D_1}$ reduces to $1/e$. At the same time, $p_S$ raises to $40.7\%$, and $p_{P_1}=11.8\%$, $p_{P_2}=6.7\%$, $p_{others}=1.3\%$.
Finally, at $T=12 ~{\rm ms}$, when atoms at $\Gamma$ point in $\rm D_1$ band almost completely decay, the proportion of atoms in S band reaches $55.8\%$, which is nearly four times the number that in the second highest band $\rm P_1$.

In order to connect the scattering channels with the proportion of atoms in each band, we add the scattering channels to the same band and get the theoretical cross section to each band.
Taking the cross section of S band as an example, the cross section is equal to twice the proportion of the channel $ss$ plus the proportion of the channel $sp_1$, $sp_2$, $sd_1$, and so on.
Fig. \ref{fig:fig5} (a) shows the final proportions of different bands obtained in experiments (orange bars) and the normalized theoretical cross sections(green bars).
The theoretical proportion of atoms scattering to S band is $57.3\%$, which is roughly consistent with the experiment point $55.8\%$. Further, the proportion of atoms in other bands is much lower than that in S band both in experiment and theory indicating that the scattering channel $ss$ is indeed dominant. Besides, we attribute the higher population of $\rm P_1$ ($\rm P_2$) band in the experiment than that in the theory to the background in the absorption imaging and the two or more collisions in the evolution time.

Further, we study the influence of optical lattice depth $V_0$ on the dominant scattering channel. Fig.\ref{fig:fig5}(b) demonstrates the experimental and theoretical proportions of final proportion of atoms in the S band at different lattice depth $V_0$. For the lattice depth $V_{0}=3,5,8,10 ~{\rm E_r}$, the experimental measurements (orange squares in Fig.\ref{fig:fig5}(b)) are $55.8\%, ~56.2\%,~58.3\%,~62.6\%$, which are close to the theoretical points (green diamonds) $57.3\%,~61.0\%,~58.3\%,~55.3\%$, respectively. The dominant scattering channels $ss$ always exists with different lattice depth, which is roughly consistent with theoretical calculation in Fig.\ref{fig:fig1.2} (a).

\section{Scattering channels in bipartite lattices}
Through the above theoretical calculation and experiments, we find that the overlap area of eigenstates between bands with different parity (for example D band and P band) in lattice with non-orthogonal lattice vectors (like triangular lattice) is much smaller than that in lattice with orthogonal lattice vectors (like square lattice). However, the orthogonality of lattice vectors has little effects on the overlapping area between bands with the same parity (for example D band and S band). Hence in the lattice with non-orthogonal lattice vector, the scattering channel between bands with the same parity will be dominant. 

For an optical lattice potential, the effect of non-orthogonal lattice vectors produces the x-y dimensional coupling term, as shown in Eq. (\ref{V}).
In order to demonstrate the relationship between the lattice geometry and the dominant scattering channels, we calculate the scattering cross sections from the $\Gamma$ point of $\rm D_1$ band in other lattices with or without x-y dimensional coupling term. In the following calculation, the first seven bands are considered and lattice depth is $V_{0}=5 ~{\rm E_r}$.

\begin{table}[htp]
\caption{The proportion of scattering cross section in lattice with different geometric structure with $V_{0}=5 ~{\rm E_r}$.}
\centering
\label{table1}
\begin{tabular}{p{2.2cm}p{1.2cm}p{1.2cm}p{1.2cm}p{1.8cm}}
\hline
\specialrule{0em}{1pt}{1pt}
{Lattice}    & {$\rm \bar{S}$ band}    & {$\rm \bar{P}$ band}    & {$D_1$ band}  & {$R_D$} \\ 
{~}    & {(average)}    & {(average)}    & {~}   \\
\hline
\specialrule{0em}{1pt}{1pt}
{1D lattice} & {$0.174$}    & {$0.349$}    & {$0.300$} & {$0.50$}   \\
{Triangular lattice} & {$0.610$}    & {$0.065$}    & {$0.142$} & {$4.30$} \\
{Square lattice} & {$0.209$}    & {$0.207$}    & {$0.187$}  & {$1.01$}  \\
{Bipartite square lattice} & {$0.382$}    & {$0.059$}    & {$0.050$} & {$6.47$}   \\
{Honeycomb lattice} & {$0.317$}    & {$0.092$}    & {$0.052$} & {$3.45$}  \\
\hline
\end{tabular}
\end{table}
	
Table. \ref{table1} shows the normalized cross section in different lattice. The proportion for $\rm \bar{S}$ band ($\rm \bar{P}$ band) represents the average proportion for each S band (P band). The proportion for $\rm D_1$ band reflects the atoms scattering to the $\rm D_1$ band. The $R_D$ is defined as:
\begin{eqnarray}
	\label{RD}
	R_D={\rm \bar{S}}/max({\rm \bar{P}},{\rm D_1}).
\end{eqnarray}
$R_D$ denotes the ratio of cross section between S band and the second biggest band, which indicates whether there exists a dominant scattering channel.
	
Without x-y dimensional coupling term, the cross section in 1D lattice to S band, P band, and D band are similar and the ratio $R_D=0.50$.
Hence, there is no dominant channel, which is also demonstrated in our recent work \cite{arXiv:2104.08794}.
	
Square lattice and triangular lattice have been discussed in Sec. II. The lattice vectors of triangular lattice are non-orthogonal and those of square lattice are orthogonal. The $R_D$ for square lattice is $1.01$, which indicates there is no dominant scattering channel. In the triangular lattice, the $R_D$ is $4.30$, which shows the cross section to $S$ band is dominant.
	
Bipartite square lattice and honeycomb lattice are both bipartite lattices, which are easily achieved in experiments by changing the polarizations of beams \cite{Jin_2019}. The potential of bipartite lattice is composed of two sets of simple lattice, and the potential of the bipartite square lattice is written as:
	\begin{eqnarray}
	\label{CSL}
	&&V=V_0 \cos(\vec{k_x}\cdot \vec{r})+V_0 \cos(\vec{k_y}\cdot \vec{r}) \\ \nn
	&&+V_0 \cos((\vec{k_x}+\vec{k_y})\cdot \vec{r})+V_0 \cos((\vec{k_x}-\vec{k_y})\cdot \vec{r}). 
	\end{eqnarray}
The third and fourth terms of the potential $V_0 \cos((\vec{k_x}+\vec{k_y})\cdot \vec{r})+V_0 \cos((\vec{k_x}-\vec{k_y})\cdot \vec{r})$ are x-y dimensional coupling term. Similarly, honeycomb lattice also has x-y dimensional coupling term. 
As shown in Table \ref{table1}, for these two types of lattices, over $30\%$ atoms scatter to each S band (there are two S bands), while only a few atoms jump to P bands. The $R_D$ of bipartite square lattice is $6.47$, and that of honeycomb lattice is $3.45$, which are both much larger than 1. It indicates that the scattering cross sections to S bands are dominant in these two types of lattices.

From the calculation, we demonstrate that the scattering channels in lattice without x-y dimensional coupling term have no significant difference. However, for lattice with x-y dimensional coupling term, the overlapping areas between bands with different parity are greatly reduced. For example, the ratio of channel $p_1p_1$ (ratio of absolute strength) in triangular lattice to that in square lattice is $0.29$, while the ratio of channel $ss$ is $0.75$. Hence, in lattice with x-y dimensional coupling terms, the decrease of scattering cross section between energy bands with different parity symmetry is more than that between bands with the same parity symmetry, which induces the channel between the same parity symmetry dominant. 
	
\section{Conclusion}
	
In conclusion, our experimental measurements and theoretical calculation unveil a dominant scattering channel in the excited bands of the triangular optical lattice. After the atoms evolve in the $\rm D_1$ band of triangular lattice for a certain time, the proportion of atom scattering to S band reaches $55.8\%$, which is around four times the proportion of the second largest band. 
Our further analysis of the different configuration lattices demonstrates that x-y dimensional coupling term is the key factor for the dominant scattering channel. This work demonstrates scattering cross sections in the 2D optical lattice and paves the way to investigate the collisions in optical lattices, which is important for control of atoms in excited bands. Furthermore, the dominant scattering channel contributes to achieving directional enhancement.

\section{Acknowledgement}
We thank Peng Zhang, Shina Tan and Xiaopeng Li for helpful discussion. This work is supported by the National Natural Science Foundation of China (Grants No. 61727819, No. 11934002, No. 91736208, and No. 11920101004) and the National Basic Research Program of  China (Grant No. 2016YFA0301501).
	
\appendix
\addcontentsline{toc}{section}{Appendices}\markboth{APPENDICES}{}
\begin{subappendices}
		
\section{Calculation of Scattering Cross Section}
We consider the two-body $s$-wave collision, and use scattering theory to calculate the scattering cross section. 
For $^{87}$Rb, the higher order of d-wave scattering can be ignored when temperature is below $200 ~{\rm \mu K}$ \cite{RN1697}. Hence the $s$-wave approximation is reasonable for BEC around $80 ~{\rm nK}$ in our experiment.

The incident wave packet can be written:
		\begin{eqnarray}
		\label{initstate}
		&&\left | \Psi_{a,b} \right \rangle =\int {\rm d}k_z {\rm d}\vec{q}_1 {\rm d}\vec{q}_2 \exp(-i(a\hat{e}_a+b\hat{e}_b)\cdot \vec{K})\\ \nn
		&&\times \phi(\Lambda)\left | \Lambda, d,d \right \rangle,
		\end{eqnarray}
where $\hat{e}_a$ and $\hat{e}_b$ are unit vectors of the cross section. $\Lambda=(k_z,\vec{q}_1,\vec{q}_2)$ and $\vec{K}=((\vec{q}_1-\vec{q}_2)/2,k_z)$ are composed of the momentum $k_z$ and quasi momentum $\vec{q}_1$, $\vec{q}_2$ of two states. $a$ and $b$ can choose any integer. $\phi(\Lambda)$ is a wave packet, which peaks at $\Lambda$. The $\left | \Lambda, n_1,n_2 \right \rangle$ are eigenstates of two atoms as:
		\begin{eqnarray}
		\label{eigstate}
		&&\left | \Lambda, n_1,n_2 \right \rangle =\exp(ik_zz+i\vec{q}_1\cdot\vec{r}_1+i\vec{q}_2\cdot\vec{r}_2)\\ \nn
		&&\times u_{n_1,\vec{q}_1}(\vec{r}_1)u_{n_2,\vec{q}_2}(\vec{r}_2),
		\end{eqnarray}
where $u_{n_1,\vec{q}_1}(\vec{r}_1)$, $u_{n_2,\vec{q}_2}(\vec{r}_2)$ are eigenstates of a single atom on $n_1,~n_2$ band, and are calculated by secular equation.

According to scattering theory, the scattering cross section of atoms jumping to band $n_1$ and $n_2$ is \cite{PhysRevA.87.063638}:
		\begin{eqnarray}
		\label{csection}
		&&\sigma(n_1,n_2)=\left | \hat{e}_a \right | \left | \hat{e}_b \right | \sum_{a,b}\int {\rm d}k'_z {\rm d}\vec{q}'_1 {\rm d}\vec{q}'_2 \\ \nn
		&&\times \left | \left \langle  \Lambda', n_1,n_2|\hat{S}-1|\Psi_{a,b} \right \rangle \right |^2,
		\end{eqnarray}
where $\Lambda'$ is parameter of the final state of scattering, and $\hat{S}$ is the scattering operator. 
Using the Born approximation the $\hat{S}$ can be calculated:
		\begin{eqnarray}
		\label{calll}
		&&\left \langle  \Lambda', n_1,n_2|\hat{S}-1|\Psi_{a,b} \right \rangle= \\ \nn
		&&-2\pi i \delta(E_{\Lambda', n_1,n_2}-E_{\Psi_{a,b}}) \left \langle  \Lambda', n_1,n_2|U(\vec{r})|\Psi_{a,b} \right \rangle,
		\end{eqnarray}
where 
		\begin{eqnarray}
		\label{Urrr}
		U(\vec{r})=\frac{4\pi a_s}{m}\delta(\vec{r})\frac{\partial}{\partial r}(\vec{r}\cdot)
		\end{eqnarray}
is interaction operator. For the $F=2,m_F=+2$, $^{87}$Rb atoms, the $s$-wave scattering length $a_s=90 ~a_B$, where $a_B$ is Bohr radius.
We defined the overlapping integral $\zeta_{n_1 ,n_2}(\vec{q}_1,\vec{q}_2;\vec{q}'_1,\vec{q}'_2)$ as:
		\begin{eqnarray}
		\label{overint}
		&&\zeta_{n_1 ,n_2}(\vec{q}_1,\vec{q}_2;\vec{q}'_1,\vec{q}'_2)=\int_{\vec{r}} {\rm d} \vec{r} \\ \nn
		&&\times u^*_{n_1,\vec{q}'_1}(\vec{r}) u^*_{n_2,\vec{q}'_2}(\vec{r}) u_{d,\vec{q}_1}(\vec{r})u_{d,\vec{q}_2}(\vec{r}),
		\end{eqnarray}
where $\vec{q}_1,\vec{q}_2,\vec{q}'_1,\vec{q}'_2$ are the quasi momentum of the initial and final states of two atoms. 
		
Using the overlapping integral (\ref{overint}), Eq. (\ref{calll}) is simplified as:
		\begin{eqnarray}
		\label{calll2}
		&&\left \langle  \Lambda', n_1,n_2|\hat{S}-1|\Psi_{a,b} \right \rangle \\ \nn
		&&\propto \zeta_{n_1 ,n_2}(\vec{q}_1,\vec{q}_2;\vec{q}'_1,\vec{q}'_2),
		\end{eqnarray}

In our experiment, atoms initially distribute at $\Gamma$ point of $\rm D_1$ band ($\vec{q}_1=\vec{q}_2=0$). The system obeys the conservation of momentum, and it causes $\vec{q}'_1=-\vec{q}'_2=\vec{q}$.
		Hence, the scattering cross section
		\begin{eqnarray}
		\label{csectionfff}
		\sigma(n_1,n_2)\propto \int {\rm d}\vec{q} ~|\zeta_{n_1 ,n_2}(0,0;\vec{q},-\vec{q})|^2.
		\end{eqnarray}
We use the periodic boundary approximation, and reduce the right side of the above formula to $\int {\rm d}\vec{q} \int d\vec{r}\times |u^*_{n_1,\vec{q}}(\vec{r}) u^*_{n_2,-\vec{q}}(\vec{r}) u_{d,0}(\vec{r}) u_{d,0}(\vec{r})|^2$. By calculating the overlapping integral, we can get the proportion of each scattering channel and the differential scattering cross section, as shown in Fig. \ref{fig:fig1.2} (a) and (b).
		
\section{Shortcut Sequences}
Shortcut is a robust method to load BEC from harmonic trap into optical lattice. In our previous work \cite{Zhou_2018,PhysRevLett.121.265301,PhysRevA.87.063638}, we have used the method to load atoms into S band and higher bands of 1D or 2D lattice. The basic principle of shortcut is that the evolution operators $U_{\vec{k}}(t)$ of momentum states $\left |\vec{k} \right \rangle$ are different between the lattice on and off. As shown in Fig.\ref{fig:fig1_new} (c), after several laser pluses, the final state of atoms is:
		\begin{eqnarray}
		\label{state_aftertwopluse}
		\left | \psi_f \right \rangle = \sum_{\vec{k}} \prod \limits_{i=n}^1 U^{off}_{\vec{k}}(t^{\rm off}_{i}) U^{on}_{\vec{k}}(t^{\rm on}_i) \times \left |\vec{k} \right \rangle,
		\end{eqnarray}
where $n$ is the number of pulses, $U^{on/off}_{\vec{k}}(t)$ is the evolution operator when lattice is on/off.
\begin{table}[htp]
\caption{The shortcut sequences to load atoms into the $\Gamma$ point of the $\rm D_1$ band in triangular optical lattice with different lattice depth.}
\label{tab:initial_seq}
\begin{tabular}{cccccccccc}
\hline
\specialrule{0em}{1pt}{1pt}
{$V_0$}    & {$t^{\rm on}_1$}    & {$t^{\rm off}_1$}    & {$t^{\rm on}_2$}    & {$t^{\rm off}_2$}    & {$t^{\rm on}_3$}  & {$t^{\rm off}_3$}  & {$t^{\rm on}_4$}  & {$t^{\rm off}_4(\mu s)$}  & {Fidelity}  \\
\hline
\specialrule{0em}{1pt}{1pt}
{$3 E_r$} & {$13.5$}    & {$11.5$}    & {$49.0$}    & {$9.5$}    & {$8.5$}  & {$56.5$}  & {$11.0$}  & {$11.0$}  & {$0.9995$} \\
{$5 E_r$} & {$29.5$}    & {$15.5$}    & {$16.5$}    & {$29.5$}    & {$6.5$}  & {$31.0$}  & {$18.0$}  & {$12.5$}  & {$0.996$} \\
{$8 E_r$} & {$10.5$}    & {$24.5$}    & {$18.5$}    & {$13.0$}    & {$10.5$}  & {$55.5$}  & {$12.5$}  & {$10.0$}  & {$0.993$} \\
{$10 E_r$} & {$59.5$}    & {$2.5$}    & {$20.5$}    & {$33.5$}    & {$12.0$}  & {$13.0$}  & {$13.5$}  & {$6.0$}  & {$0.975$} \\
\hline
\end{tabular}	
\end{table}
		
By choosing the number of pulses and pulse length, we can optimize the final state $\left | \psi_f \right \rangle$ to aimed state $\left | \psi_a \right \rangle$. The fidelity is defined by $|\left \langle \psi_f|\psi_a \right \rangle |^2$ to describe the loading efficiency. In the experiment, the optimized sequence has four pluses to load atoms into $\rm D_1$ band of triangular optical lattice, and the pulse sequences of different lattice depth are shown in Table \ref{tab:initial_seq}.

\end{subappendices}

\bibliographystyle{apsrev}
\bibliography{my}

\end{document}